# Bragg-Williams order competes with superconductivity

Xu Liu[1,2,†], Xu Chen[1,†], Chuizhen Chen[3,†], Boqin Song[1], Jing Chen[1], Xijing Dai[1], Qinghua Zhang[1], Feng Jin[1], Xingya Wang[4], Weiwei Dong[5], Dongliang Yang[5], Gefei Li[1], Pengju Zhang[1], Jiangping Hu[1], Jiangang Guo[1], Tianping Ying[1,*], Xiaolong Chen[1,2,*]

1. Beijing National Laboratory for Condensed Matter Physics, Institute of Physics, Chinese Academy of Sciences, Beijing 100190, China.
2. School of Physical Sciences, University of Chinese Academy of Sciences, Beijing 100049, China.
3. School of Physical Science and Technology, Soochow University, Suzhou 215006, China
4. Shanghai Synchrotron Radiation Facility, Shanghai Advanced Research Institute, Chinese Academy of Sciences, Shanghai 201204, China
5. High Energy Photon Source, Institute of High Energy Physics, Chinese Academy of Sciences, Beijing 101407, China
* ying@iphy.ac.cn
* xlchen@iphy.ac.cn

**ABSTRACT**

Orderings in charge and spin have been extensively studied to unravel their correlation to emergent superconductivity over the past decades. Bragg-Williams order (BWO), a classical structural order parameter describing site occupancy in alloys, has long been speculated to influence superconducting behavior. Yet, its role still remains ambiguous, largely due to the difficulty of isolating BWO from concomitant charge doping or competing electronic instabilities. Here, we establish $In_{2/3}PSe_3$ as a platform wherein indium vacancies are reversibly configurable between ordered and disordered states via thermal treatment. We show that the disordered phase undergoes a pressure-induced superconducting transition with a $T_c$ of 11 K, significantly higher than the 7 K observed in its ordered counterpart. This constitutes a rare instance in which pure BWO variation drives a substantial shift in $T_c$. By combining a Ginzburg–Landau phenomenological analysis with a BCS–McMillan microscopic description, we demonstrate that BWO naturally suppresses superconductivity through electron-phonon interactions, a mechanism supported by ultra-low-wavenumber Raman measurements. Our findings support BWO as an independent order parameter that competes directly with superconductivity, extending the concept of competing orders beyond conventional electronic and magnetic degrees of freedom.

**Main**

Crystalline imperfections such as mixed occupancies and vacancies are ubiquitous companions in superconductors, yet their direct role in the superconductivity remains poorly understood[1-9]. A prime example is found in the $YBa_2Cu_3O_{7-x}$, where the order-disorder transition of oxygen in the Cu-O chains governs both the tetragonal-to-orthorhombic structural transition and the emergence of superconductivity[10,11]. A similar situation arises in the iron-based superconductor $K_xFe_{2-y}Se_2$, where ordering of iron vacancies nucleates a robust antiferromagnetic insulating state that competes with the superconducting phase[12–15]. In these systems, however, altering the atomic configuration inevitably changes the chemical composition, introducing excess oxygen or iron as charge doners. As a result, the superconducting phase diagram has been almost universally framed in terms of carrier concentration, with superconductivity competing against charge or spin degrees of freedom including charge density wave (CDW), spin density wave (SDW) or antiferromagnetism[16,17].

Apart from the charge and spin order mentioned above, a well-defined structural order, Bragg-Williams order (BWO), has been widely used in alloys and intermetallics[18]. In its simplest schematic form, BWO is illustrated by the order parameter measuring the excess of correctly occupied over incorrectly occupied sites with respect to a chosen ordered reference pattern,

$$\eta = \frac{R - W}{R + W},$$

where $R$ and $W$ denote the number of right and wrong occupied atomic positions, respectively (Fig. 1a). By construction, $\eta = 0$ corresponds to a fully random configuration, while $\eta = 1$ represents perfect order. The thermodynamic consequence of BWO is a competition between Coulomb repulsive energy ($E$) and configurational entropy ($S$). To illustrate this generic order-disorder physics, we consider a simplified square lattice with 50% site occupancy, for which the $E$ and $S$ can be expressed as functions of $\eta$ (see Supplementary Material for detailed derivation):

$$E = \frac{1}{2}U(1 - \eta^2)$$

$$S = k_B\{ln2 - \frac{1}{2}[(1 + \eta)\ln(1 + \eta) + (1 - \eta)\ln(1 - \eta)]\}$$

The resulting Helmholtz free energy ($F = E - TS$) is plotted numerically in Fig. 1b. A critical temperature $T_c$ emerges, below which the ordered state ($\eta \neq 0$) is thermodynamically stable, while above $T_c$ the disordered state ($\eta = 0$) minimizes the free energy. This temperature-dependent behavior implies that by quenching the material from different thermal histories below $T_c$, one can selectively tune $\eta$ to a desired value[11].

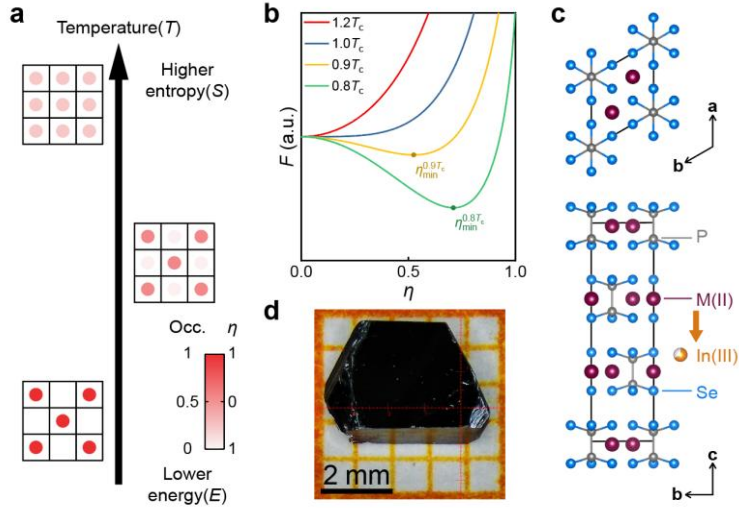

FIG 1. **a**, Schematic of a simplified two-dimensional square model with 50% of atomic occupation, illustrating how atomic configurations evolve with the BWO parameter $\eta$, and the resulting competition between repulse energy and configurational entropy. **b**, Helmholtz free energy as a function of $\eta$ at different temperatures. **c**, Crystal structure of MPSe$_3$ and design strategy. **d**, Optical image of In$_{2/3}$PSe$_3$ single crystal.

To isolate BWO from carrier change and transfer, we therefore begin with a nonmagnetic insulator or semiconductor that exhibits an order-disorder transition, thereby minimizing electronic contributions and maximizing the influence of BWO. The $M$PSe$_3$ family ($M$ = metal) provides an ideal starting point: these layered materials possess semiconducting ground states that suppresses free carriers[19,20], exhibit a propensity for pressure-induced superconductivity[21–25], and offer the flexibility to choose $M$ from across the periodic table[19,20]. A key insight is that charge balance constrains *M* to be divalent. Therefore, selecting a trivalent metal, such as $In^{3+}$, naturally introduces vacancies into the system[26,27]. This leads us to a novel vacancy-ordered triclinic $In_{2/3}PSe_3$ (O-phase, $\eta = 1$), and its gradual transformation to a vacancy-disordered rhombohedral phase (D-phase, $\eta = 0$) by tuning the quench temperature.

**Discovery of O-phase $In_{2/3}PSe_3$ and its tunable $\eta$**

The $M$PSe$_3$ family provides an ideal framework within which $\eta$ can be continuously tuned. As showed in Fig. 1c, the structure consists of a framework of P and Se atoms forming distorted octahedra, with interstices occupied by metal ions with an average valence of +2. Single crystalline $In_{2/3}PSe_3$ were grown using iodine-assisted chemical vapor transport. As shown in Fig. 1d, the resulting crystals exhibit trigonal symmetry, inherited from the trigonal $[P_2Se_6]^{4-}$ framework. Crucially, the O- and D-phases can be reversibly switched through annealing and quenching. Energy dispersive X-ray spectroscopy (EDS) and XRD confirm that both phases share identical composition and lattice parameters (Supplementary Fig. S2), demonstrating that thermal history alone dictates $\eta$ without altering stoichiometry.

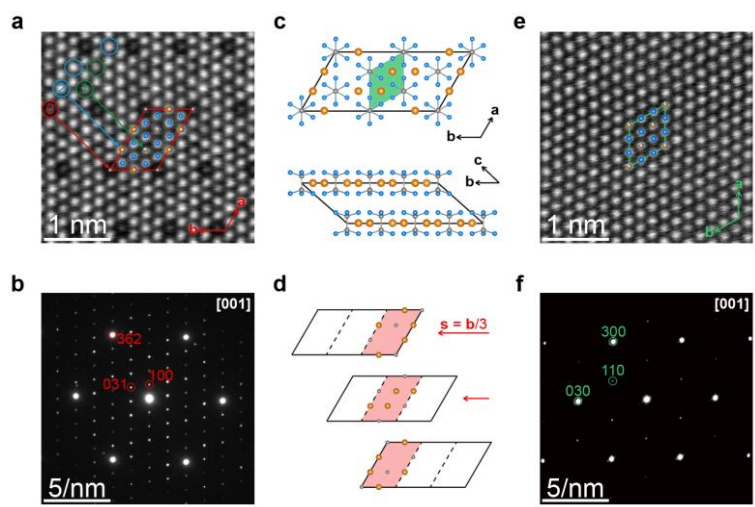

FIG 2. **a**, HAADF image and **b**, indexed SAED image of O-phase. **c**, Crystal structure of the O-phase. **d**, Schematic of the trilayer ordered phase stacking, whit only the In and P atoms within the highlighted region are shown. **e**, HAADF image and **f**, indexed SAED image of the D-phase.

To directly visualize the vacancy ordering, we performed aberration-corrected scanning transmission electron microscopy (STEM) on both phases. The high-angle annular dark-field (HAADF) images of the O-phase taken along the in-plane direction (Supplementary Fig. S3) are fully consistent with the D-phase, revealing a characteristic interlayer shift of 1/3. However, out-of-plane HAADF images of the O-phase (Fig. 2a) reveal a symmetry distinct from the expected *R*-3 structure, exhibiting an in-plane superstructure with periodicity of $2\sqrt{3}/3 \times \sqrt{3}$. This superstructure is corroborated by the corresponding selected-area electron diffraction (SAED) patterns (Fig. 2b). Accounting for the 1/3 interlayer shift, the actual monolayer periodicity expands by a factor of three, resulting a final superlattice period of $2\sqrt{3} \times \sqrt{3}$. Close examination of the contrast in Fig. 2a reveals three distinct intensity levels within each periodic unit in addition to the Se atom, marked by blue, green, and red circles. Each position contains one site hosting a P-P dimer and two sites occupied by indium atoms. The three contrast levels therefore correspond systematically to indium atom counts of 2, 1, and 0 per site, respectively. This intensity modulation is fully consistent with an average indium site occupancy of 2/3, confirming the ordered vacancy arrangement in the O-phase. We systematically enumerate in Supplementary Fig. S4 all configurations that yield the projected intensity pattern of Fig. 2a. Of these, two are centrosymmetric, and the remaining one lack inversion symmetry.

The absence of a second harmonic generation (SHG) signal in the O-phase (Supplementary Fig. S5), confirms the preservation of inversion symmetry. First-principles calculations identify configuration 1 as the ground state (Supplementary Table S1). This result aligns with the electrostatic intuitive that interionic repulsion favors maximally separation of vacancies. Figure 2c presents the refined crystal structure of the O-phase, which adopts space group $P$-1 and corresponds to an order parameter $\eta = 1$. Figure 2d illustrates how layer stacking gives rise to the observed HAADF pattern. For clarity, only 1/3 of each layer are shown, capturing the interlayer slip vector $s = b/3$, while selenium atoms are omitted and each P-P dimer is represented by a single phosphorus atom.

Figures 2e and 2f present HAADF and SAED patterns of the D-phase, revealing an absence of superstructure reconstruction. The SAED patterns can be well indexed to the space group $R$-3, isostructural with other members of the $M$PX$_3$ family[22]. The D-phase therefore adopts $R$-3 symmetry, with indium atoms exhibiting fully random site occupancy, corresponding to an order parameter $\eta = 0$. Direct comparison of the SAED patterns between the two phases reveals that the diffraction spots of the D-phase align perfectly with the fundamental reflections of the O-phase. This observation confirms that their structural distinction originates solely from the indium-site dictated BWO, while the underlying framework remains unchanged.

**Pressure-induced superconductivity in O- and D-phases**

Physical pressure offers a clean tuning knob that modifies the lattice without altering the chemical composition. With the crystal structures of both O- and D-phases established, we turned to high-pressure electrical transport measurements. Figures 3a and 3b show that the resistance of O-phase decreases gradually with increasing pressure. Superconductivity emerges at approximately 27 GPa with an onset transition temperature ($T_c^{onset}$) of 3 K. Upon further compression, $T_c$ rises monotonically, reaching 7 K by the highest measured pressure. The D-phase follows a similar trajectory but with striking differences. As shown in Fig. 3c and 3d, superconductivity appears at a lower pressure of 21 GPa, but a much higher $T_c^{onset}$. With increasing pressure, $T_c$ continues to climb, eventually saturating near 11 K. Notably, this value represent the highest $T_c$ reported to date for any superconductor in the $M$PX$_3$ family.

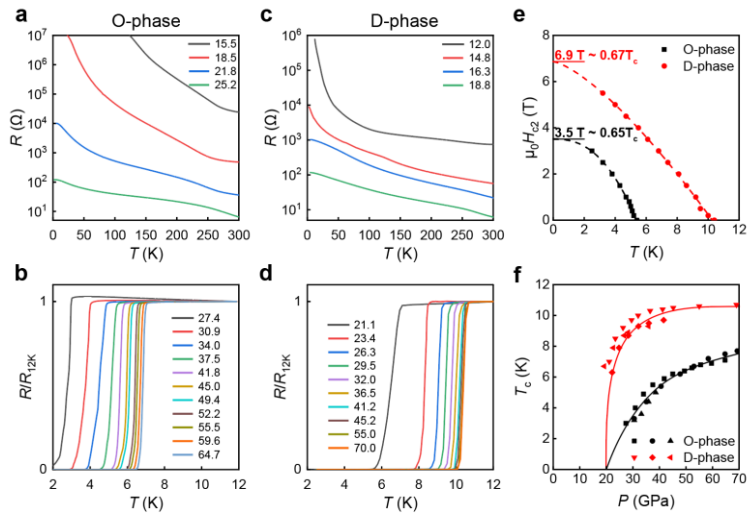

FIG 3. Temperature-dependent resistance of the **a-b**, O- and **c-d**, D-phases of In$_{2/3}$PSe$_3$ with the variation of external pressures. **e**, The temperature dependence of the upper critical field with the fitting at 37 GPa.

**f**, Pressure-dependent superconducting transition temperature for the O- and D-phases of In$_{2/3}$PSe$_3$.

Figure S6 displays the electrical resistance of O- and D-phases as a function of temperature at various magnetic field measured at approximately 37 GPa. The observed shift of the resistivity transition toward lower temperatures and its broadening under magnetic fields further confirm the observed superconductivity. With $T_c$ defined as the temperature at 90% of the normal-state resistivity, we extracted the relationship between the $T_c$ and the upper critical field ($\mu_0 H_{c2}$) for the two phases, as shown in Fig. 3e. The data are well described by the generalized Ginzburg-Landau (G-L) formula across the accessible temperature range. Notably, the extracted values of $\mu_0 H_{c2}(0)/T_c$ are nearly identical, being 0.65 T/K and 0.67 T/K for the O- and D-phases, respectively. This close agreement indicates that the underlying pairing mechanism is the same in both phases, despite their different vacancy configurations. Figure 3f summarizes the pressure evolution of $T_c$ for both phases.

We further performed high-pressure synchrotron X-ray diffraction measurements on both phases. Figure 4a and 4b display the color contour of the diffraction evolution for both phases. Upon compression, the diffraction peaks gradually shift to higher angles and progressively broaden, eventually leading to amorphization at the highest pressures as other pressurized $M$PX$_3$ compounds[25,28]. Notably, no evidence of a structural phase transition is detected, even as the sample enters the superconducting state. The unit cell volume also decreases smoothly with increasing pressure, showing no anomalies or discontinuities (Supplementary Fig. S8).

As superconductivity inevitably introduces charge carriers at the Fermi level, distinguishing the contribution of the density of states from that of vacancy order is essential. We measured the Hall resistivity of both phases (Supplementary Fig. S9) and extracted the corresponding carrier concentrations, plotted in Fig. 4c. A quantitative comparison is revealing. At 30 GPa, the D-phase already achieves a $T_c$ of 10 K, significantly higher than the maximum $T_c$ of the O-phase (7 K), which occurs only at 70 GPa where the carrier concentration is an order of magnitude larger. Given the absence of magnetic elements in our framework, pressure-induced changes in carrier density cannot account for the enhanced superconductivity in the D-phase. Instead, the dominance of BWO emerges as the controlling factor. Therefore, our observation hints a pure manifestation of BWO competing with superconductivity, as illustrated in Fig. 4d. We note that the degree of disorder is tunable, and a partially ordered state is also presented in Fig. S10.

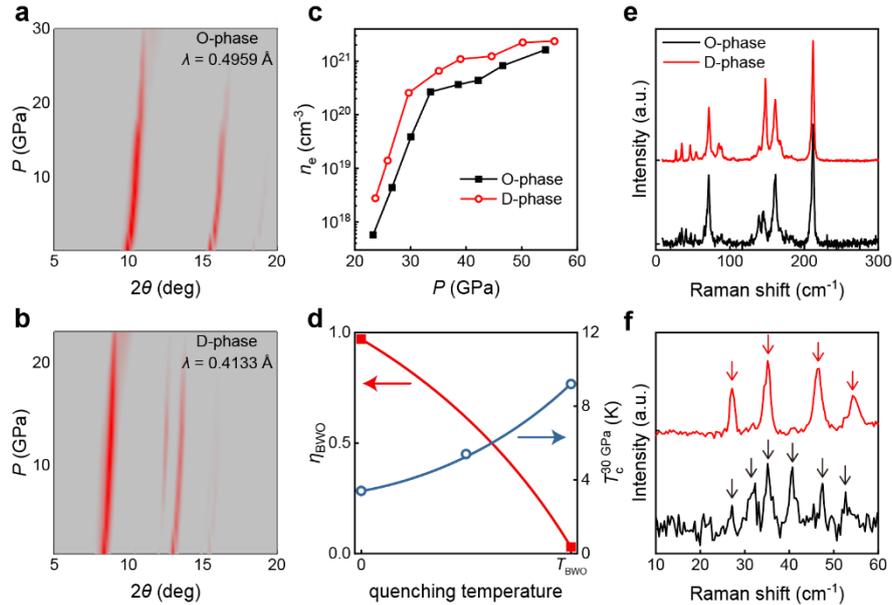

FIG 4. **a** and **b**, Synchrotron powder X-ray diffraction patterns for the O- and D-phases under pressure. Patterns were measured at separate beamlines with different wavelengths. **c**, Carrier concentration as a function of pressure for the O- and D-phases. **d**, Phase diagram of the BWO and superconductivity in In$_{2/3}$PSe$_3$. **e** and **f**, Raman spectra of the O- and D-phases.

**Phenomenological and microscopic interpretation**

To elucidate the interplay between BWO and superconductivity, we formulate a phenomenological Ginzburg-Landau theory[29]. In the absence of lattice order ($\eta = 0$), the system exhibits an intrinsic critical temperature $T_{c0}$. The competition between the phases is captured by a repulsive coupling term, ($\gamma \mid \Delta \mid^2 \mid \eta \mid^2$), between the superconducting order parameter $\Delta$ and the BWO parameter $\eta$. This coupling renormalizes the quadratic coefficient of the superconducting phase, $\alpha_0(T - T_{c0})$, where $\alpha_0$ is a positive constant, directly yielding a suppressed transition temperature:

$$T_c = T_{c0} - \frac{\gamma}{\alpha_0} \mid \eta \mid^2,$$

where $\gamma > 0$ represents the coupling constant between the two order parameters. This macroscopic relation inherently captures our central observation: the ordered O-phase ($\eta = 1$) exhibits a significantly lower $T_c$ relative to the disordered D-phase ($\eta = 0$).

The microscopic origin of this repulsive coupling lies in the lattice dynamics governed by the indium-vacancy configuration. By treating the indium sublattice as a lattice gas, the structural degrees of freedom are defined by local density fluctuations relative to the stoichiometric background, $\delta n_i = n_i - 2/3$. In the O-phase, interionic Coulomb repulsion establishes a long-range vacancy correlation field, $C(R) = \langle \delta n_i \delta n_{i+R} \rangle$. for which $\eta$ serves as the macroscopic amplitude.

Crucially, these periodic positional correlations perturb the local force constants. Taking the ensemble average over vacancy configurations, the BWO enhances the stiffness of pairing-relevant phonon modes. For a specific phonon branch $v$ at momentum $q$, the frequencies shift according to:

$$\omega_{qv,\eta}^2 = \omega_{qv,0}^2 + \eta^2 \delta \omega_{qv}^2,$$

where $\eta^2 \delta \omega_{qv}^2 > 0$ for the vacancy-sensitive modes. This disordering in D-phase ($\eta = 0$) manifests as a softening of the lattice: $\omega_{qv}^2$ decreases monotonically as $\eta$ is reduced from 1 to 0[30,31].

Within the frameworks of Migdal-Eliashberg and BCS-McMillan theories, the dimensionless electron-phonon coupling constant $\lambda$ is inversely proportional to the mean-square phonon frequency[32–34]. The stiffening of the phonon network in the ordered state therefore suppresses this coupling to leading order:

$$\lambda(\eta) \approx \lambda_D - \lambda_2 \eta^2,$$

where $\lambda_2 > 0$. Because $T_c$ scales exponentially with $-1/\lambda(\eta)$, the superconducting transition temperature is monotonically reduced as vacancy order increases.

**Phonon fingerprints of BWO**

Raman phonon spectroscopy provides direct evidence that vacancy ordering modifies the lattice dynamics. Figure 4e compares the Raman spectra of the O-phase and D-phase over a broad frequency range. The three dominant phonon peaks located at approximately 72 cm$^{-1}$, 162 cm$^{-1}$, and 210 cm$^{-1}$ are nearly identical in both phases. These modes arise primarily from the rigid P-Se framework, which remains largely unchanged upon vacancy disordering. Above 50 cm$^{-1}$, however, subtle but clear differences emerge. The peaks around 85 cm$^{-1}$ and 145 cm$^{-1}$ exhibit noticeable intensity variations between the two phases, indicating a systematic redistribution of spectral weight from higher to lower frequencies.

The most striking difference appears below 50 cm$^{-1}$. As shown in Fig. 4f, the O-phase displays significantly more vibrational peaks than the D-phase. This is a direct consequence of the lower crystallographic symmetry of the O-phase (*P*-1) compared with the D-phase (*R*-3). The reduced symmetry and expanded supercell activate additional Raman-active modes, which could be attributed to the heavier indium sublattice. Admittedly, Raman spectroscopy only probes phonons at the Brillouin zone center, which cannot directly provide the evolution of $\omega_{qv}^2$. A full determination of the phonon softening across the entire Brillouin zone requires further experiments such as inelastic X-ray or neutron scattering. Nevertheless, the Raman data establish a direct link between BWO and lattice dynamics. In this context, BWO acts as a "vacancy density wave" whose primary competitive mechanism is the stiffening of the local bonding network. Conversely, the loss of long-range vacancy order in the D-phase disrupts positional correlations, softens the lattice, and profoundly enhances the electron-phonon interaction to promote a robust high-temperature superconducting state. Our results therefore identify BWO as an independent structural order parameter capable of controlling superconductivity, distinct from carrier doping, magnetism, or conventional electronic ordering.

More broadly, as a fundamental descriptor, BWO may influence a host of other collective phenomena including magnetism, polarization, optical response, each potentially tunable through the same thermal history controlled $\eta$. Our work establishes BWO as a versatile knob for engineering quantum materials, far from a secondary perturbation, can itself serve as a primary tuning parameter.

**ACKNOWLEDGMENT**

We are grateful for the fruitful discussion with Dr. Tongxu Yu. This work was financially supported by the National Key Research and Development Program of China and National Natural Science Foundation of China (Grants No. 52522201, 2021YFA1401800, 52272267, 2024YFA1611303 and 52302010). This work was supported by the Synergetic Extreme Condition User Facility (SECUF, https://cstr.cn/31123.02.SECUF).

Xu Liu[1,2,†], Xu Chen[1,†], Chuizhen Chen[3,†], Boqin Song[1], Jing Chen[1], Xijing Dai[1], Qinghua Zhang[1], Feng Jin[1], Xingya Wang[4], Weiwei Dong[5], Dongliang Yang[5], Gefei Li[1], Pengju Zhang[1], Jiangping Hu[1], Jian-gang Guo[1], Tianping Ying[1,*], Xiaolong Chen[1,2,*]
1. Beijing National Laboratory for Condensed Matter Physics, Institute of Physics, Chinese Academy of Sciences, Beijing 100190, China.
2. School of Physical Sciences, University of Chinese Academy of Sciences, Beijing 100049, China.
3. School of Physical Science and Technology, Soochow University, Suzhou 215006, China
4. Shanghai Synchrotron Radiation Facility, Shanghai Advanced Research Institute, Chinese Academy of Sciences, Shanghai 201204, China
5. High Energy Photon Source, Institute of High Energy Physics, Chinese Academy of Sciences, Beijing 101407, China
* ying@iphy.ac.cn
* xlchen@iphy.ac.cn

**1. Theoretical analysis of the order-disorder phase transition**

Qualitatively, the order-disorder phase transition originates from the competition between energy ($E$, favoring order) and entropy ($S$, favoring disorder), with temperature influencing the outcome of this competition and thus determining the degree of order in the system. Quantitatively, the relationship between the degree of order and temperature can be determined from the condition that the system tends to minimize its Helmholtz free energy ($F = E - TS$). The energy expression is determined by the specific structure of the ordered phase. To simplify the model, we consider only the nearest-neighbor interactions of the relevant atoms. The entropy expression is given by the sum of the configurational entropy of each site:

$$S = \sum_{\text{site},i} -kx_i \ln x_i,$$

where $x_i$ represents the occupancy of atom $i$ at the site. Next, we will perform thermodynamic analysis and present numerical solutions for the classical solid solution model as well as the ordered phase structure in this work.

I. Solid solution model with equal numbers of A and B atoms

This model contains two types of sites with equal numbers, and the atomic occupancies are given by:

$$\text{site 1}: Occ_\text{A} = \frac{1}{2}(1+\eta), Occ_\text{B} = \frac{1}{2}(1-\eta)$$

$$\text{site 2}: Occ_\text{A} = \frac{1}{2}(1-\eta), Occ_\text{B} = \frac{1}{2}(1+\eta),$$

where $\eta$ represents the order parameter in the system, degree of order. This formula also serves as the definition of the degree of order in this model. Since the two sites are related by atomic exchange, they share the same entropy, which simplifies the calculation. Therefore, the average energy and entropy per site can be expressed in terms of the order parameter:

$$E = \frac{Z}{2}V_\text{AB} + \frac{ZU}{4}(1-\eta^2)$$

$$S = k\{ln2 - \frac{1}{2}[(1+\eta)\ln(1+\eta) + (1-\eta)\ln(1-\eta)]\}$$

Where $Z$ represents the nearest-neighbor coordination number, $V$ represents the interatomic interaction energy, and $U = \frac{1}{2}(V_{AA} + V_{BB}) - V_{AB}$. The formation of a solid solution requires $U > 0$; otherwise, segregation will occur. The expression for the Helmholtz free energy can then be obtained:

$$F = E - TS$$
$$= \frac{Z}{2}V_{AB} + \frac{ZU}{4}(1-\eta^2) - kT\{ln2 - \frac{1}{2}[(1+\eta)\ln(1+\eta) + (1-\eta)\ln(1-\eta)]\}$$
$$= \frac{ZU}{4}\{\frac{T}{T_c}[(1+\eta)\ln(1+\eta) + (1-\eta)\ln(1-\eta)] - \eta^2\} + F(0),$$

where $T_c = \frac{ZU}{2k}$, $F(0) = \frac{Z}{2}V_{AB} - kTln2 + \frac{ZU}{4}$. The minimum point of the Helmholtz free energy $F$ corresponds to the degree of order $\eta$ of the system. Examining the expression for the Helmholtz free energy reveals that the degree of order depends only on temperature. The corresponding results are shown in Fig S1a and S1b. The order parameter as a function of temperature exhibits characteristics of a second-order phase transition, similar to the order-disorder transition in CuZn alloys.

II. In$_{2/3}$PSe$_3$ model

This model contains two types of sites in a 2:1 ratio, with the atomic occupancies given by:

$$\text{site 1: } Occ = \frac{2}{3} + \frac{1}{3}\eta$$

$$\text{site 2: } Occ = \frac{2}{3} - \frac{2}{3}\eta,$$

where $\eta$ represents the order parameter in the system, degree of order. This formula also serves as the definition of the degree of order in this model. Therefore, the average energy and entropy per site can be expressed in terms of the order parameter:

$$E = \frac{U}{6}(4 - \eta^2)$$

$$S = k\{ln3 - [\frac{2}{9}(2+\eta)\ln(2+\eta) + \frac{2}{9}(1-\eta)\ln(1-\eta) + \frac{1}{9}(2-2\eta)\ln(2-2\eta) + \frac{1}{9}(1+2\eta)\ln(1+2\eta)]\},$$

where $U$ represents the interaction energy between indium atoms. The expression for the Helmholtz free energy can then be obtained:

$$F = \frac{U}{6}(4-\eta^2) - kT\{ln3$$
$$- [\frac{2}{9}(2+\eta)\ln(2+\eta) + \frac{2}{9}(1-\eta)\ln(1-\eta)$$
$$+ \frac{1}{9}(2-2\eta)\ln(2-2\eta) + \frac{1}{9}(1+2\eta)\ln(1+2\eta)]\}$$
$$\approx \frac{U}{6}\{\frac{1.0223T}{T_c}[\frac{4}{9}(2+\eta)\ln(2+\eta) + \frac{4}{9}(1-\eta)\ln(1-\eta) + \frac{2}{9}(2-2\eta)\ln(2-2\eta) + \frac{2}{9}(1+$$

$$2\eta)\ln(1+2\eta)\big] - \eta^2\} + constant,$$

where $T_c \approx \frac{1.0223U}{3k}, constant = \frac{2}{3}U - kT\ln3$. As with the previous model, the degree of order depends only on temperature. However, the change in the structure of the ordered phase leads to a modification in the expression for the Helmholtz free energy. The corresponding results are shown in Fig S1c and S1d. Unlike the previous model, the order parameter undergoes a sudden change at the critical temperature and then continues to increase as the temperature decreases, exhibiting characteristics of a first-order phase transition, similar to the order-disorder transition in $Cu_3Zn$ alloys.

Through thermodynamic analysis of these two models, it can be inferred that the relationship between the degree of order and temperature is determined by the specific structure of the ordered phase. Furthermore, considering additional interactions also leads to different outcomes. However, a definitive conclusion is that temperature modulates the competition between energy and entropy, resulting in a continuous tunability of the degree of order below the critical temperature, where lower temperatures yield a higher degree of order until saturation is reached.

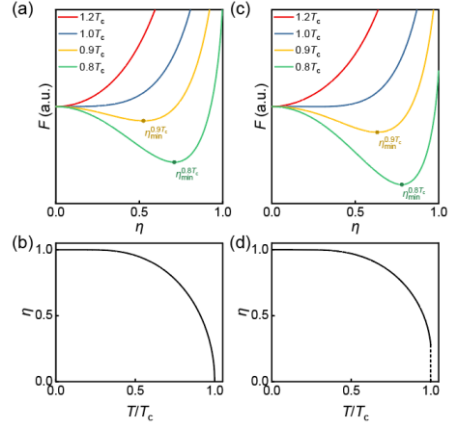

FIG S1. The Helmholtz free energy as a function of order parameter $\eta$ at different temperatures for the (a) model I and (c) model II. The normalized temperature dependence of order parameter $\eta$ for the (b) model I and (d) model II.

## 2. Crystal synthesis and characterization under ambient conditions

The order phase $In_{2/3}PSe_3$ were grown by the chemical vapor transport (CVT) method using iodine as the transport agent. Indium grains, Phosphorus lumps, and Selenium grains were weighed according to the stoichiometric ratio and sealed in a quartz tube together with iodine grains. The quartz tube was placed in a two-zone furnace, with the source zone and growth zone temperatures set at 520°C and 450°C, respectively. After one week of growth, single crystals with typical dimensions of 4×4×1 mm$^3$ were obtained at the growth end. The disordered phase $In_{2/3}PSe_3$ can be obtained by annealing the ordered phase at 600°C followed by quenching in ice water.

Energy Dispersive X-ray Spectroscopy (EDS) was performed using a Phenom ProX scanning electron microscope with electron energy of 15 keV. X-ray diffraction (XRD) was performed using a Rigaku Smart Lab diffractometer with Cu $K_\alpha$ radiation ($\lambda$ = 1.5418 Å). The high-angle annular dark-field (HAADF) and selected area electron diffraction (SAED) images were obtained using a JEOL ARM-200F scanning transmission electron microscope (STEM) operating at 200 kV. For the in-plane direction images, the sample was prepared using the focused ion beam (FIB) method.

Second Harmonic Generation was performed using an incident light with a wavelength of 1064 nm. Raman measurements were carried out using a laser with an excitation wavelength of 633 nm.

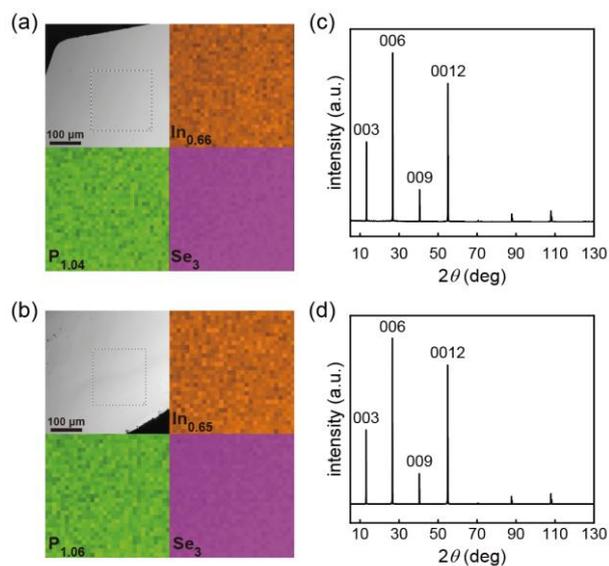

FIG S2. EDS mappings of the (a) O- and (b) D-phases. XRD patterns of the single crystals of (c) O- and (d) D-phases.

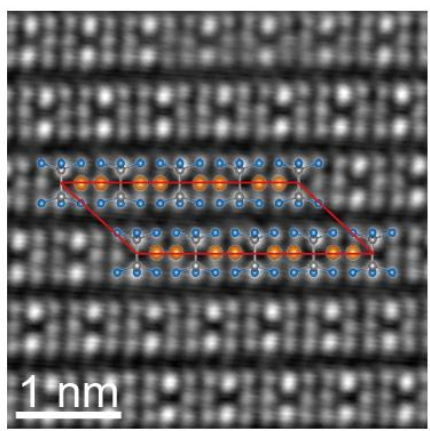

FIG S3. HAADF image of the O-phase taken along the in-plane direction.

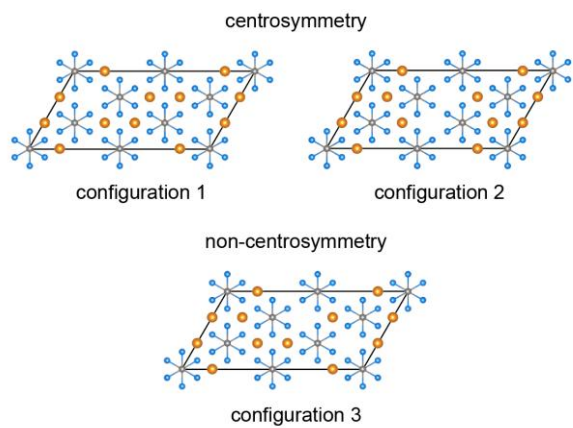

FIG S4. Possible In atom occupancies in the O-phase.

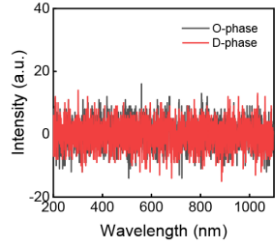

FIG S5. The SHG spectra of the O- and D-phases

## 3. First-principles calculations

First-principles calculations were carried out with the density functional theory (DFT) implemented in the Vienna ab initio simulation package (VASP)[1,2]. The generalized gradient approximation (GGA) in the form of Perdew-Burke-Ernzerhof (PBE)[3] was adopted for the exchange-correlation potentials. We used the projector augmented-wave (PAW)[4] pseudopotentials with a plane wave energy of 500 eV; $5s^25p^1$, $3s^23p^3$ and $4s^24p^4$ were treated as valence electrons for In, P and Se, respectively. A Monkhorst-Pack Brillouin zone sampling grid[5] with a resolution of $0.025\times2\pi$ Å$^{-1}$ was applied. The self-consistent field procedure was considered convergent when the energy difference between two consecutive cycles was lower than $10^{-8}$ eV. Atomic positions and lattice parameters were fully relaxed till all the forces on the ions were less than $10^{-2}$ eV/Å.

TABLE S1. Calculation results for the three structures after relaxation

|  | Stru 1 | Stru 2 | Stru 3 |
| --- | --- | --- | --- |
| S. G. | $P$-1 | $P$-1 | $P$-1 |
| $a$ (Å) | 11.17 | 11.23 | 11.21 |
| $b$ (Å) | 22.33 | 22.40 | 22.41 |
| $c$ (Å) | 10.04 | 10.05 | 10.05 |
| $\alpha$ (°) | 42.22 | 41.85 | 41.97 |
| $\beta$ (°) | 111.72 | 110.12 | 111.39 |
| $\gamma$ (°) | 120.05 | 119.21 | 120.157 |
| $V$ (Å$^3$) | 1455.66 | 1471.38 | 1459.09 |
| $E$ (eV) | -237.30 | -234.68 | -236.13 |

## 4. Electrical and structural characterization under high pressure

The high-pressure experiments were conducted using a diamond anvil cell (DAC). Diamond culet sizes of 300 μm were employed, with rhenium gaskets serving as the pressure-confining material. Pressure was calibrated using the ruby fluorescence method[6]. For electrical transport measurements, c-BN was used as the insulating material and KBr served as the pressure transmitting medium. Platinum foils were used as electrodes, and the electrical transport measurements were conducted in a commercial Physical Property Measurement System (Quantum Design). For synchrotron measurements, glycerol was used as the pressure transmitting medium. The synchrotron radiation experiments were performed at the BL17UM station of the Shanghai Synchrotron Radiation Facility (SSRF) with $\lambda$ = 0.4959 Å and Beijing High Energy Photon Source (HEPS) with $\lambda$ = 0.4133 Å. The pressure-dependent volume compression data was fitted by the Birch–Murnaghan (BM) equation of state[7,8]:

$$P = \frac{3}{2}B_0\left[\left(\frac{V_0}{V}\right)^{\frac{7}{3}} - \left(\frac{V_0}{V}\right)^{\frac{5}{3}} \times \left\{1 + \frac{4}{3}(B' - 4)\left[\left(\frac{V_0}{V}\right)^{\frac{3}{2}} - 1\right]\right\}\right]$$

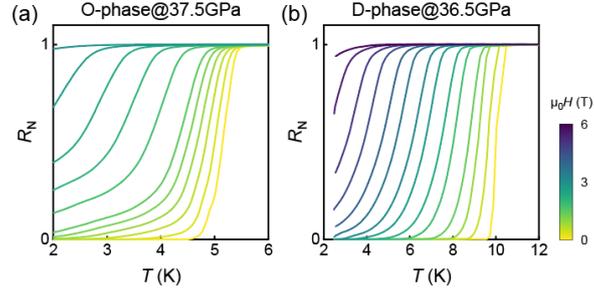

FIG S6. The superconducting transition of the O- and D-phases under different magnetic fields at around 37 GPa.

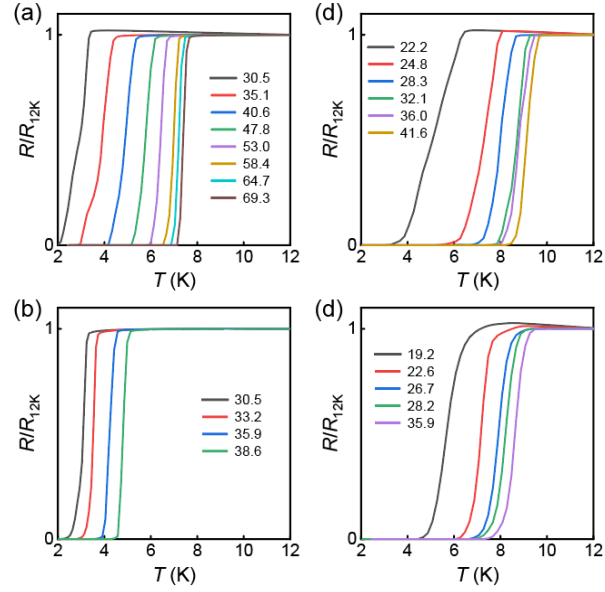

FIG S7. Temperature-dependent resistance of the (a)-(b) O-phase and (c)-(d) D-phase under different pressures.

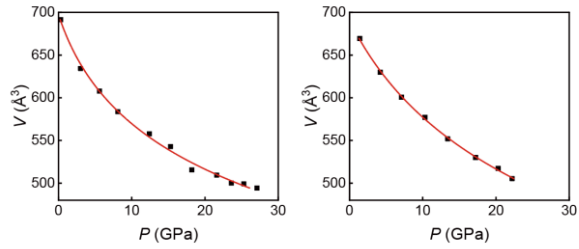

FIG S8. Pressure dependence of the unit cell volume for the (a) O- and (b) D-phases, with the red lines representing the Birch–Murnaghan fitting.

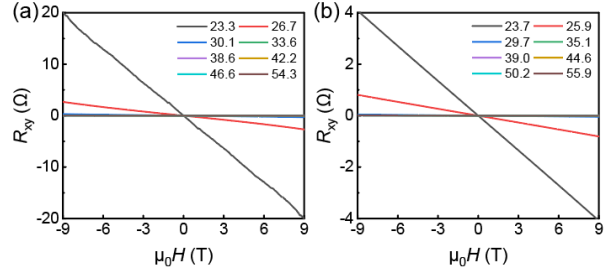

FIG S9. Hall resistance of the (a) O- and (b) D-phases.

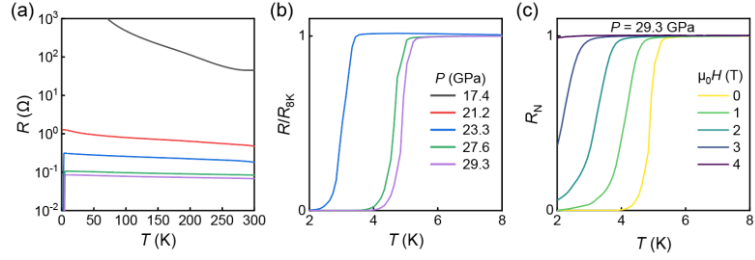

FIG S10. Temperature-dependent resistance of the partially ordered state.